\documentclass[preprint,12pt,nonatbib]{elsarticle}

\usepackage[numbers,sort&compress]{natbib}

\usepackage{amssymb}
\usepackage{amsmath}

\usepackage{lineno}
\usepackage{graphicx}
\graphicspath{{figures/}}

\begin{document}

\begin{frontmatter}



\title{Performance assessment of helicon wave heating and current drive in EXL-50 spherical torus plasmas}


\author[1]{G. J. Qiao}
\author[2,3]{D. Luo}
\author[2,3]{S. D. Song}
\author[2,3]{J. Q. Dong}
\author[2,3]{Y. J. Shi}
\author[4]{J. C. Li}
\author[1]{D. Du}
\author[2,3]{Y. K. Martin Peng}
\author[2,3]{M. S. Liu}
\author[2,3]{EXL-50 team}

\address[1]{University of South China, Hengyang, Hunan 421001, P. R. China}
\address[2]{Hebei Key Laboratory of Compact Fusion, Langfang 065001, P. R. China}
\address[3]{ENN Science and Technology Development Co., Ltd., Langfang 065001, P. R. China}
\address[4]{Department of Earth and Space Sciences, Southern University of Science and Technology, Shenzhen 518055, P. R. China}
\ead{jingchunli@pku.edu.cn, 9367645@qq.com}

\begin{abstract}
Analysis of helicon wave heating and current drive capability in EXL-50 spherical torus plasmas has been conducted. It is found that the driven current increases with the launched parallel refractive index $n_{||}$ and peaks around $n_{||} = 4.0$ when the frequency of the helicon wave is between 300~MHz and 380~MHz. The helicon wave current drive efficiency shows a relatively stable upward trend with increasing plasma temperature. Moreover, the driven current decreases as the plasma density increases. We also analyzed the current drive with helicon waves of 150~MHz and 170~MHz and found that the driven current at a lower frequency was lower than that at a higher frequency. A positive proportional relationship exists between the driven current and $n_{||}$. Besides, as $n_{||}$ increases, the profile of the driven current becomes wider. Finally, the effect of the scrape-off layer (SOL) region on the helicon wave current drive was also investigated.
\end{abstract}




\end{frontmatter}



\section{Introduction}
\label{sec1}
The fast waves in magnetized plasmas can be generally divided into two types. The first type of fast wave, referred to as ``fast wave'' in the following discussion, has a frequency in the ion cyclotron frequency range. It is what we usually call ``fast waves'' in tokamaks~\cite{1,2,3,4,5}. The second type of fast wave is called the ``helicon'' wave, whose frequency is much higher than the ion cyclotron frequency and close to the lower hybrid frequency. Such a second type of fast wave is called a ``whistle wave'' or ``fast wave in the lower hybrid frequency range''. The rays of the first type of fast wave are dominated by ion absorption in the ion cyclotron resonance layer.
On the other hand, helicon rays are dominated by complete electron absorption~\cite{6,7}. The propagation characteristics of helicon waves are similar to the whistle waves in space plasma, in which the group velocity is mainly in the direction of the background magnetic field. The helicon waves slowly spiral radially inwards and gradually deposits energy, exhibiting strong off-axis current-driven characteristics due to their slow radial propagation. The importance of helicon waves for current actuation was initially realized through ARIES reactor research~\cite{8}. The helicon wave current drive has not been evaluated experimentally. However, the current calculations show the effectiveness of helicon wave current drive~\cite{9,10,11}. Hence, using helicon waves to heat and drive current becomes essential for next-generation toroidal magnetic confinement devices such as tokamak and spherical tokamak~\cite{12}.

A series of numerical calculations have verified the high efficiency of helical current drive by the ray-tracing code GENRAY using the Ehst-Karney formula, the Fokker-Planck code CQL3D containing the quasi-linear effects, or the full-wave code AORSA~\cite{13}. In order to obtain the current driving ability of helicon waves in a specific configuration, R. Prater et al.~\cite{9} and S.J. Wang et al.~\cite{14} calculated the current driving ability of helicon waves in both DIII-D and KSTAR tokamaks, respectively. The GENRAY/CQL3D simulations show that helicon wave propagation is not as chaotic as fast waves, and the deposition rate of helicon waves is significantly substantial. Kinetic effects and the effect of DC electric fields were also investigated in DIII-D and EAST tokamaks~\cite{15}. Utilizing the quasi-linear code, the results in Ref.~\cite{15} show that quasi-linear effects are not negligible, which confirms the importance of nonlinear effects of RF waves injected with high wave power. Recently, C. Lau has performed a numerical study to understand better the effects of SOL turbulence in helicon wave propagation and absorption in the DIII-D tokamak~\cite{16}, C. Lau found that SOL turbulence can cause helicon wave refraction to undesirable locations and strong helicon wave absorption in the scrape-off layer (SOL), resulting in tremendously less helicon wave power deposited in the core plasma. Similarly, Li et al. studied the helicon wave in a toroidal plasma and found its different properties under high and low plasma parameters~\cite{17}.

Experimentally, high harmonic fast wave (HHFW) heating and current drive were explored in the NSTX device. Experiments utilizing HHFW achieved $f_{NI}\sim 0.65$ with $I_p = 300$~KA plasma (\#138506) having $P_{rf} = 1.4$~MW, using one-wave toroidal mode number $k_\varphi = -8~\mathrm{m}^{-1}$~\cite{17}. Recently, a deuterium H-mode discharge with a plasma current of 300~KA, an axial toroidal magnetic field of 0.55~T and a calculated non-induced plasma current fraction of 0.7 was also obtained in NSTX under 30 MHz HHFW heating at 1.4~MW~\cite{18}. It is important to note that the frequencies of HHFW in NSTX are much lower than the ``helicon waves'' or ``fast waves in the lower hybrid frequency range,'' and their properties are much more like the ``fast waves in the ion cyclotron frequency range.'' Regarding the helicons, relevant physical experiments have been conducted in the DIII-D tokamak for current drive and parameter decay. However, the detailed results of the corresponding experiments have not been officially reported~\cite{20}. Even though many efforts have been made to clarify the helicon properties, some aspects still need to be studied. For example, the capability of helicon wave current drive in low-density plasmas is still unclear. Moreover, the effect of the SOL region also needs more investigations, especially for a spherical torus with high beta plasma.

EXL-50 is a new spherical tokamak designed in 2018 and built in 2019. It has a major radius of 0.58~m. One of the critical experimental goals of EXL-50 tokamak is to assess the effectiveness of electron cyclotron current drive (ECCD) and electron cyclotron resonance heating (ECRH) in the absence of a center solenoid (CS) magnet. It has been demonstrated that the ECRH can achieve steady-state high-efficiency current drive and obtain the highest RF current drive record without a CS magnet~\cite{21,22,23}. Recently remarkable experiments have also shown plasma currents as high as  80-100~KA at line densities over $0.5\times 10^{19}~\mathrm{m}^{-3}$. Experiments have also revealed that a tremendous number of confined energetic electrons exist outside the LCFS (last-closed flux surface) of EXL-50 spherical tokamak, having solenoid-free ECRH sustained plasmas~\cite{24}.

This paper presents the results investigating the wave heating and current drive capability of EXL-50 tokamak discharges. We find that when the frequency of the helicon wave is between 300~MHz and 380~MHz, the driven current increases with the increase of $n_{||}$, and peaks around $n_{||} = 4.0$. Our result also shows that the driven current increases with the increase of plasma core temperature. In the meanwhile, the driven current first decreases and then increases with the increase of plasma core density. It is suggested that the helicon wave can drive a very considerable current at extremely low plasma density. The following discussion also gives some new results about the effect of the SOL region.

\section{Helicon wave and GENRAY code}
\label{sec2}
It is well-known that the power of helicon waves is mainly because of electron Landau damping and transit time magnetic pumping. In warm Maxwellian plasmas, the wave deposition in plasmas with electron temperature Te and electron density ne is expressed as follows.
\begin{equation}
\label{eq1}
K_{\perp i}=\frac{\sqrt{\pi}}{4} K_{\perp} \beta_e \xi_e e^{-\xi_e^2} G,
\end{equation}
$k_{\perp i}$ is the imaginary part of the perpendicular wave number $k_{\perp }$, $\xi_e = v_{||}/v_{te} = c/n_{||}v_{te}$. Moreover, $v_{te} = (2kT_e/m_e)^{1/2}$  is the electron thermal speed, $v_{||}$ is the phase velocity of the wave parallel to the equilibrium magnetic field, {\itshape c} is the speed of light, and $n_{||}$ is the parallel refraction index. The formula of $k_{\perp}$ can be found in Ref.~\cite{9,13}.

As mentioned, we use the Ray-tracing code GENRAY to calculate the helicon wave heating and current drive in the spherical tokamak. GENRAY has several options for power absorption and current driving. The Chiu~\cite{25} model is used for electron and ion absorption in which the code calculates the vertical component of the imaginary part of the refractive index. Additionally, the standard Ehst-Karney model is used for the current drive. The density and temperature profiles can be written with the following empirical formula~\cite{26}.
\begin{equation}
\label{eq2}
\begin{aligned}
n_e&=(n_{ec}-n_{ea})[1-(r/a)^i]^s+n_{ea}\\
T_e&=(T_{ec}-T_{ea})[1-(r/a)^i]^s+T_{ea}\\
\end{aligned}
\end{equation}
$n_{ea}$ and $n_{ec}$ are the edge and core electron densities, respectively. $T_{ea}$ and $T_{ec}$ are the edge and core electron temperatures. The $i = 1$ and $s = 1/2$  are the empirical indexes. We kept the temperature and density profiles constant in the calculations, and only the temperature and density change at the center of the plasma was considered.

GENRAY presently incorporates the SOL module. Calculations using the SOL model include information on magnetic fields, plasma distribution, and geometry. The magnetic field geometry in the SOL region is directly imported from the equilibrium reconstruction utilizing the EFIT code. Outside the LCFS region, the density and temperature of the plasma are given in the following equations.
\begin{equation}
\label{eq3}
\begin{aligned}
n_\rho&=n_{L C F S} \times e^{\left(-(\rho-1) / \delta_n\right)}\\
T_\rho&=T_{L C F S} \times e^{\left(-(\rho-1) / \delta_T\right)}
\end{aligned}
\end{equation}
$n_{LCFS} = n_{e,i}(a)$ is electron and ion densities at the last closed flux surface. The $T_{LCFS} = T_{e,i}(a)$ represent the temperature at the last closed flux surface. $\delta_n$ is the normalized exponential density with respect to plasma radius outside of the LCFS beginning at $\rho = 1$. Finally, the $\delta_T$ is the temperature decay lengths outside of the LCFS beginning at $\rho = 1$.

\section{Numerical results}
\label{sec3}
\subsection{Helicon wave heating and current drive}
\label{sec3.1}
\begin{figure}[!ht]
\centering
\includegraphics[width=0.86\columnwidth]{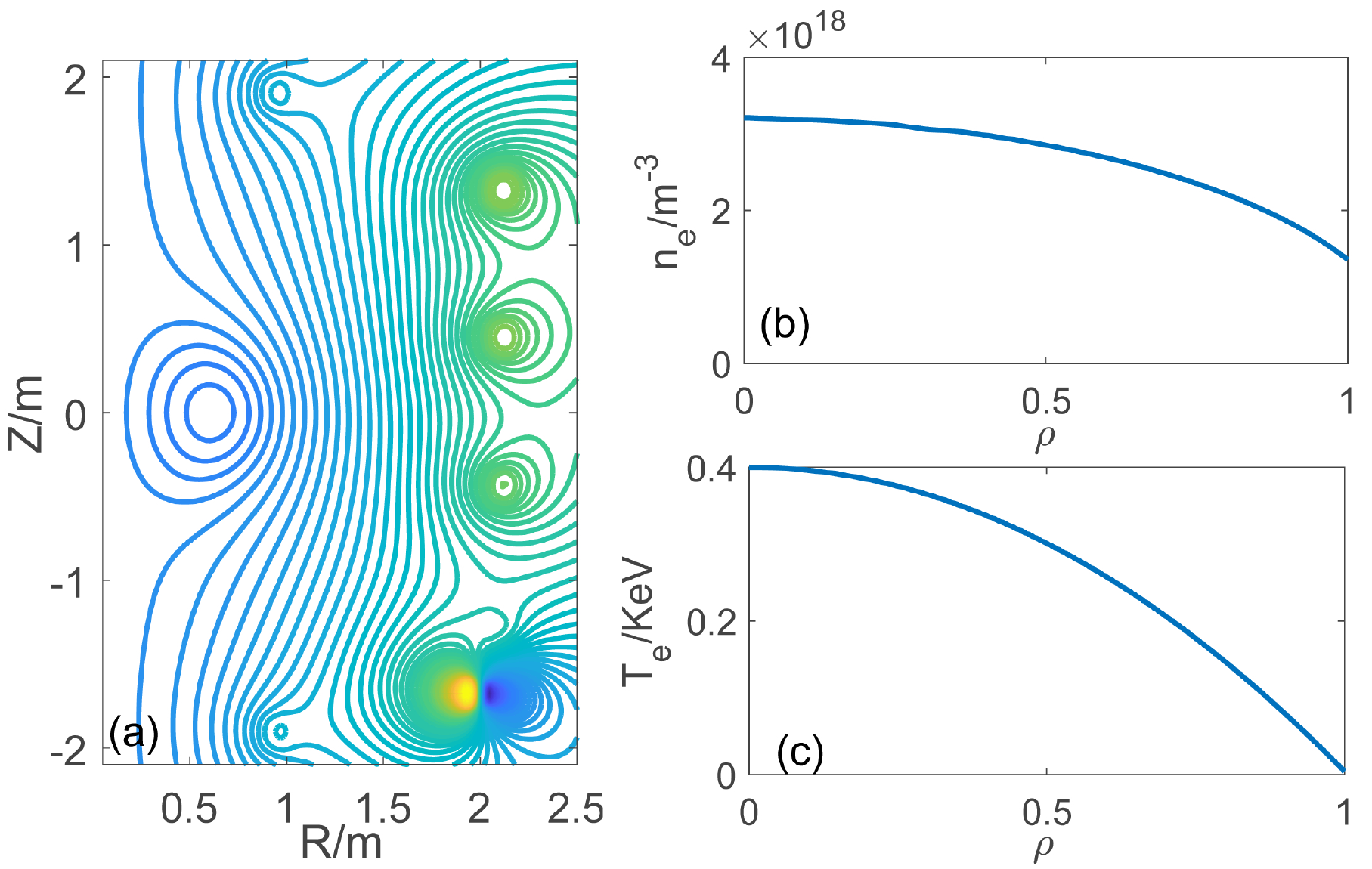}
\caption{Configuration, profiles of density and temperature}
\label{fig1}
\end{figure}

We first performed the numerical analysis of helicon heating and current drive under the equilibrium configuration of EXL-50 tokamak discharge. The equilibrium of the EXL-50 tokamak was derived from the reconstruction of the EFIT code by using the experimental \#9237 for 2.0 seconds. The equilibrium was a limited configuration. The toroidal magnetic field $B_0$ was 0.3~T, and the deuterium plasma current was 42~KA. Fig.~\ref{fig1}(a) shows the magnetic equilibrium configuration of the EXL-50 tokamak through the reconstructions of the 9237th shot. Fig.~\ref{fig1}(b,c) presents the temperature and density profiles used for the simulation. The central temperature of electrons was 400~eV, and the center density of electrons was $3.2\times 10^{18}~\mathrm{m}^{-3}$ unless otherwise stated.
\begin{figure}[!ht]
\centering
\includegraphics[width=0.68\columnwidth]{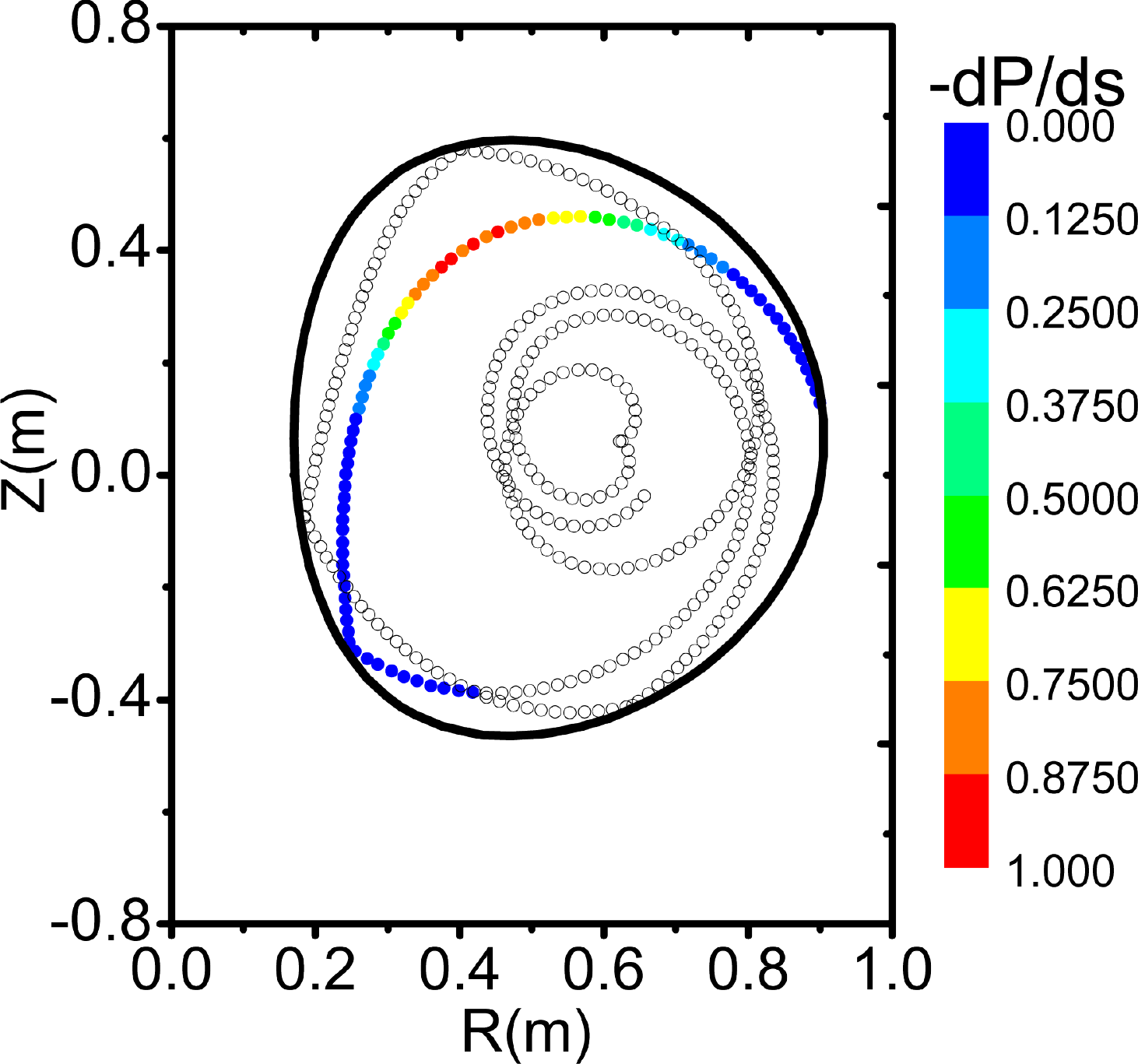}
\caption{Helicon wave simulation for a typical EXL-50 scenario}
\label{fig2}
\end{figure}

Fig.~\ref{fig2} illustrates a helicon wave propagation trajectory with a frequency of 300~MHz, a plasma center temperature of 300~eV, and a core plasma density of $0.8\times 10^{18}~\mathrm{m}^{-3}$. The center parallel refractive index $n_{||}=3.4$, and the input power is 1~MW. It can be seen that the helicon wave is absorbed quickly after entering the plasma, and the position, where the wave is absorbed, is located at the off-axis region of plasma most of the time. For this case, the total current was about 187~KA.
\clearpage

\begin{figure}[!ht]
\centering
\includegraphics[width=0.6\columnwidth]{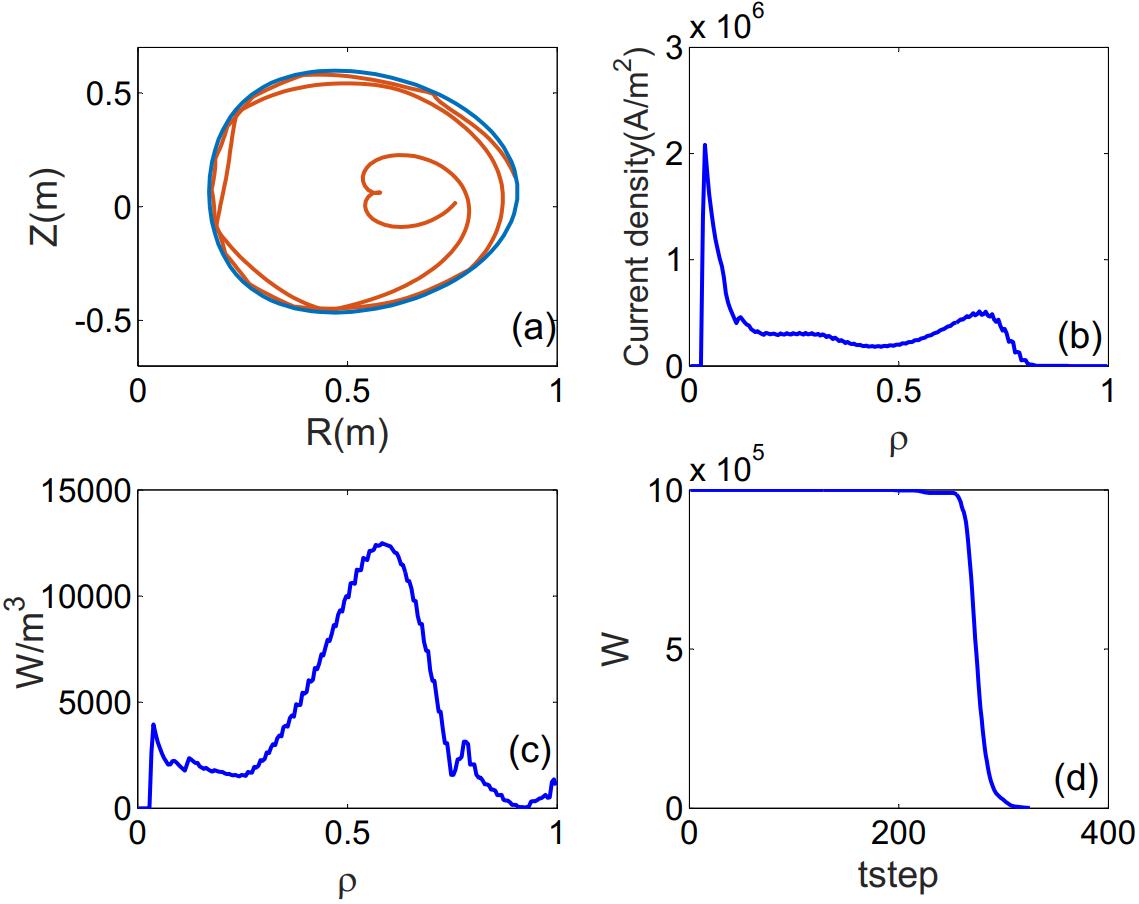}
\caption{Wave trajectories (a),  profile of driven current (b), power deposition profile (c) and residual power in the trajectories with time (d) for a 260MHz helicon wave}
\label{fig3}
\end{figure}

Fig.~\ref{fig3}(a,b,c) presents the wave trajectory of the helicon wave with $f=260$~MHz and $n_{||} = 4.4$ , the current profile driven by the helicon wave, and the power deposition distribution, respectively. Fig.~\ref{fig3}(d) shows the residual power that remained in the trajectory as a function of the time step. It can be seen that the single-pass absorption dominates for these plasma parameters. Moreover, the power absorption of the helicon wave lies in a broader off-axis plasma region.

\begin{figure}[!ht]
\centering
\includegraphics[width=0.8\columnwidth]{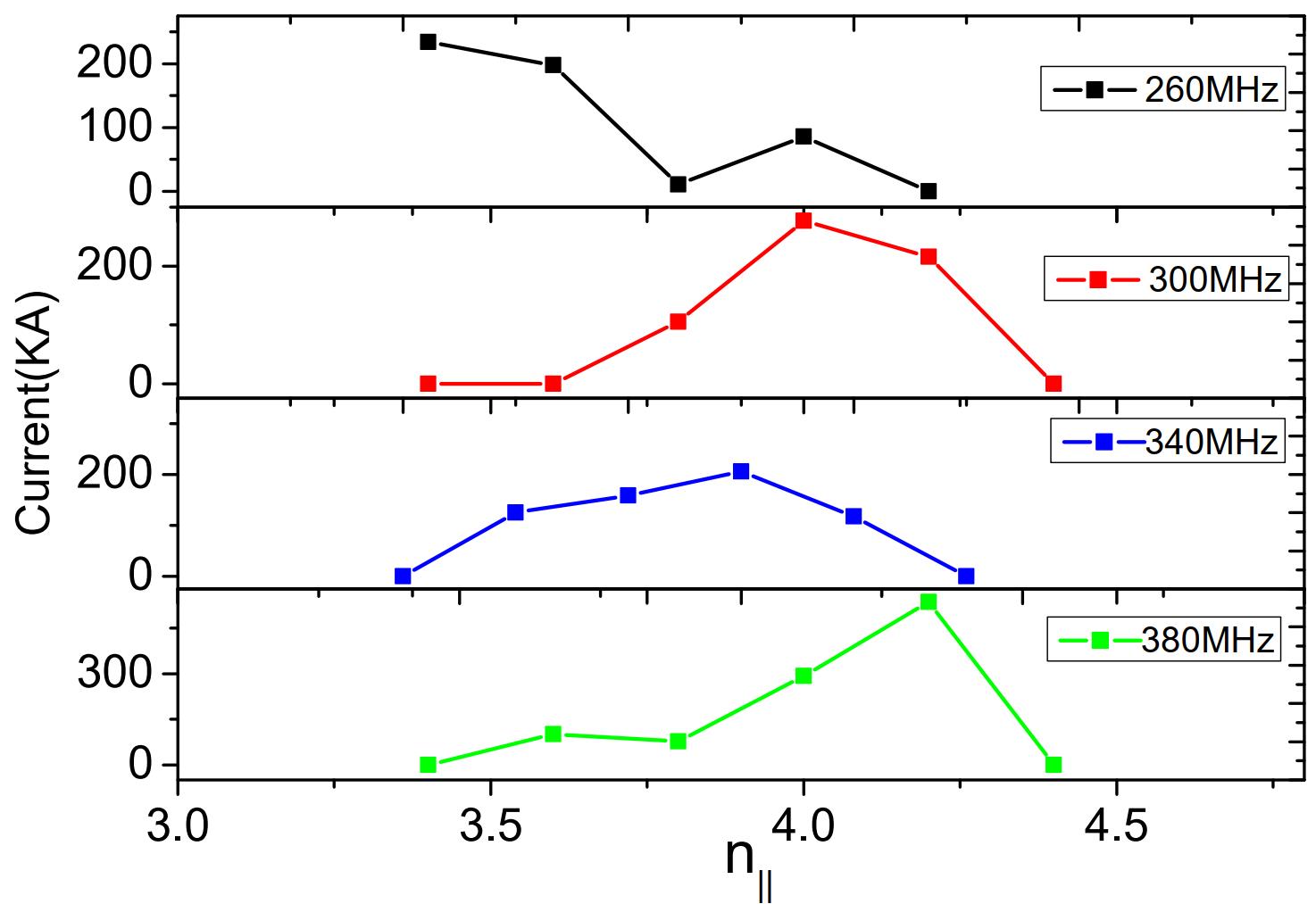}
\caption{Dependence of driven current on $n_{||}$ at different frequencies}
\label{fig4}
\end{figure}

The helicon wave current drive was recorded as a function of $n_{||}$ at different frequencies. Fig.~\ref{fig4} shows the change of driven current with $n_{||}$ for a core temperature of 400~eV and plasma density of $3.2\times 10^{18}~\mathrm{m}^{-3}$. It can be seen that when the helicon wave frequency is 300~MHz--380~MHz, the driven current increases with the increase of $n_{||}$ and peaks at around $n_{||}$ = 4.0. The driven current efficiency is sensitive to the launched $n_{||}$. Therefore, for the engineering design of the helicon wave system, the value of $n_{||}$ should be set around 4.0 to obtain a higher drive current. At the same time, it can be observed that the driven current is 0 in all three calculation cases. Hence, the value of $n_{||} = 4.4$ should be avoided in the design.

\begin{figure}[!ht]
\centering
\includegraphics[width=0.95\columnwidth]{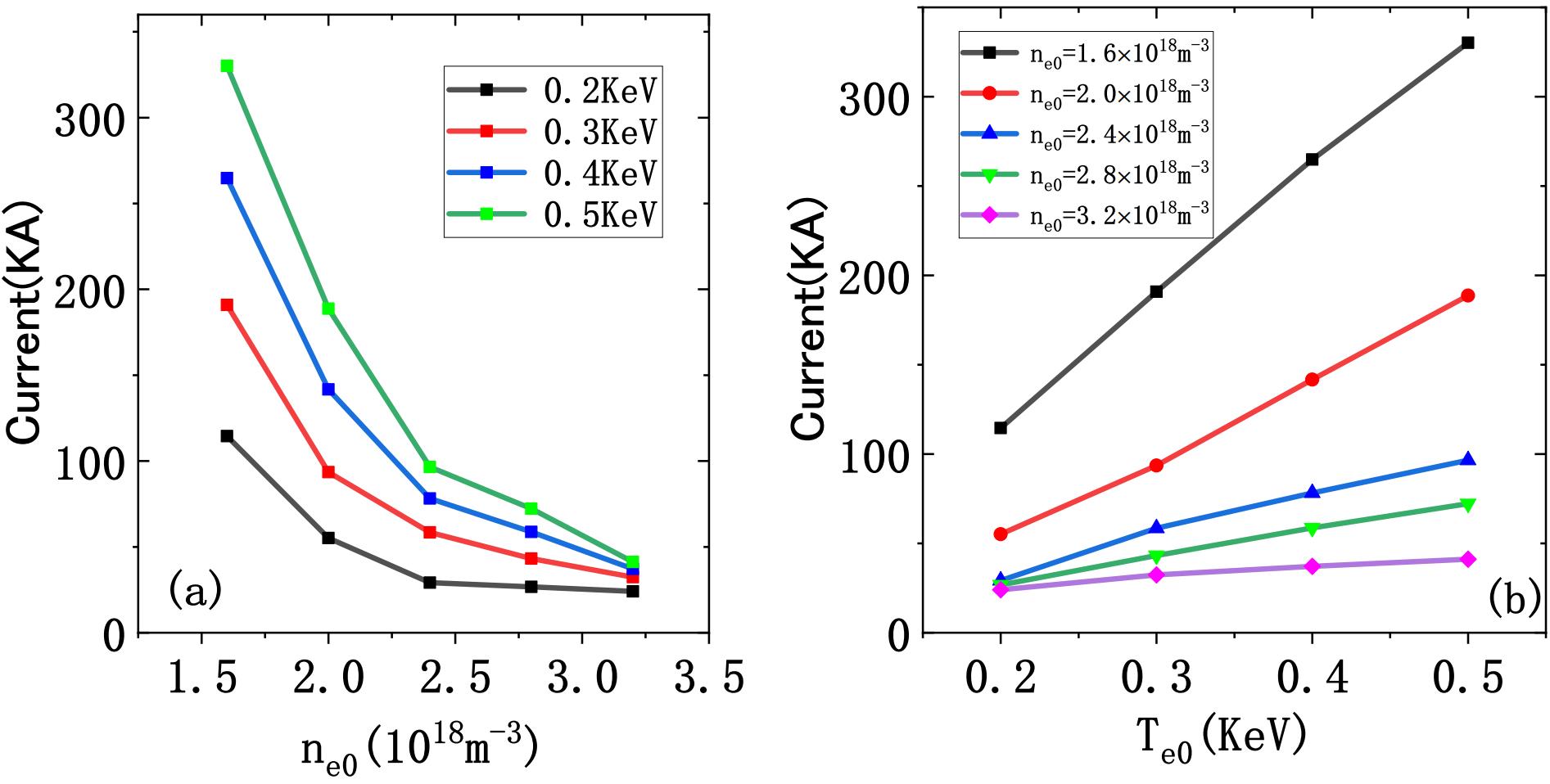}
\caption{Helicon wave (300MHz) driven current as functions of core temperature and density plasma}
\label{fig5}
\end{figure}

The core temperature and central density of the plasma can strongly affect the wave heating and current driving ability. Therefore, the effect of varying plasma temperatures and densities was studied on the wave current-driving ability of the 300~MHz helical wave. Fig.~\ref{fig5} shows the variation of the helical wave current driving at 300~MHz with different temperatures and densities. Fig.~\ref{fig5}(a) is the variation of driven current with the variation of plasma temperature for the fixed plasma density (0.8--$3.2\times 10^{18}~\mathrm{m}^{-3}$); Fig.~\ref{fig5}(b) is the change of driven current with the change of plasma density for a specific plasma temperature (0.2--5~KeV). In Fig.~\ref{fig5}(a), the driven current has a relatively stable upward trend with increased plasma temperature. The higher plasma temperature can effectively increase the driven current. As per Fig.~\ref{fig5}(b), the effect of variation in the plasma density on the driven current is regular. When the plasma density changes, the driven current always decreases in these cases.
There are two exciting results to observe. The first result is that the magnitude of the driven current does decrease monotonically with the increase of the density. The second result is that the wave at a lower plasma density drives considerable current. For example, when the density is $1.0\times 10^{18}~\mathrm{m}^{-3}$, a 1~MW helicon wave can drive a current of about 500~KA. Compared with the simulation results of the previous HL-2M and DIII-D tokamaks, this driving efficiency is quite attractive.

\subsection{Helicon wave current drive at 150~MHz and 170~MHz}
\label{sec3.2}
\begin{figure}[!ht]
\centering
\includegraphics[width=0.85\columnwidth]{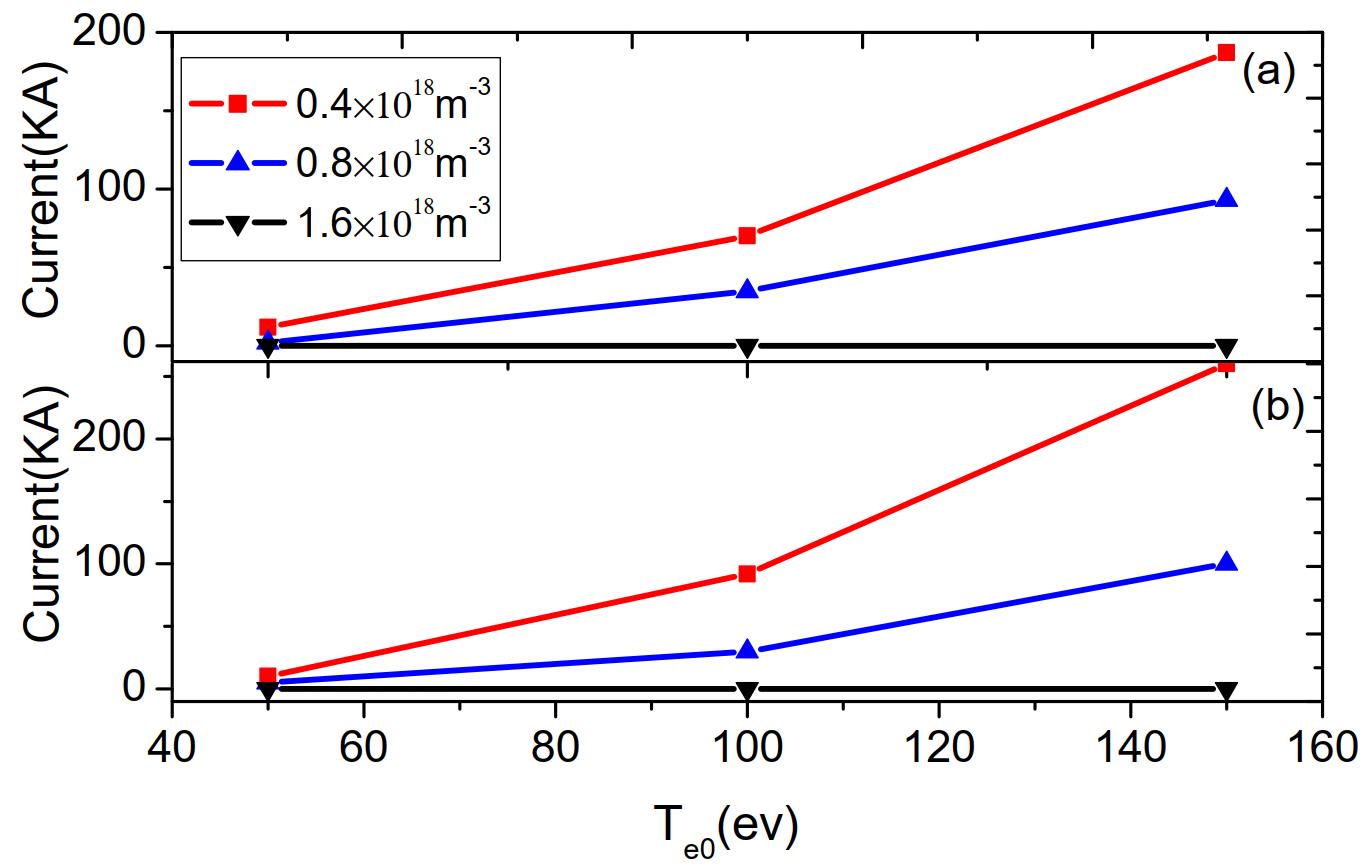}
\caption{Helicon wave of 150MHz(a) and 170MHz(b) driven current versus electron temperatures}
\label{fig6}
\end{figure}

So far, we have presented the driving capability under different wave frequencies and found that higher frequencies give a higher driven current. For instance, a current of 550~KA can be achieved with a frequency of 380~MHz. Due to the availability of klystrons, the wave frequency should be fixed before installing the wave system. To investigate the characteristics of helicon waves at a fixed frequency, we did a more detailed study on the helicon waves of 150~MHz and 170~MHz in the EXL-50 device. Fig.~\ref{fig6} shows the relationship between the change in plasma temperature and the driven current at different helicon frequencies with the plasma densities of  $0.4\times 10^{18}~\mathrm{m}^{-3}$, $0.8\times 10^{18}~\mathrm{m}^{-3}$, and $1.6\times 10^{18}~\mathrm{m}^{-3}$ at 150~MHz and 170~MHz. Combined with the results in Fig.~\ref{fig5}, it can be seen that at low frequencies, with the increase of plasma temperature, the driven current still has a relatively stable upward trend. We also note that a higher plasma density at this helicon wave frequency is unsuitable for generating driven current.
\begin{figure}[!ht]
\centering
\includegraphics[width=0.8\columnwidth]{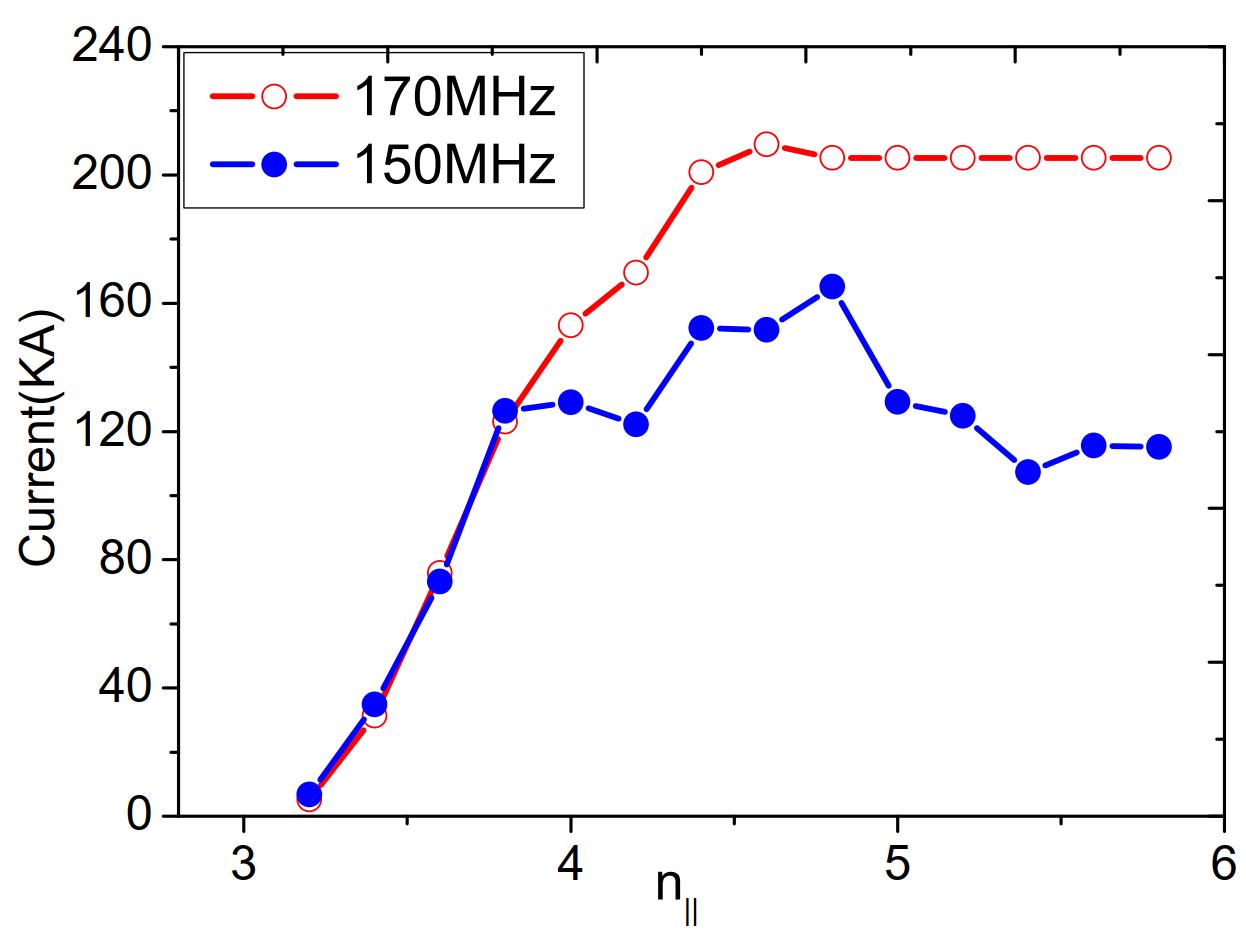}
\caption{helicon wave of 150MHz and 170MHz driven current versus $n_{||}$}
\label{fig7}
\end{figure}

\begin{figure}[!ht]
\centering
\includegraphics[width=0.9\columnwidth]{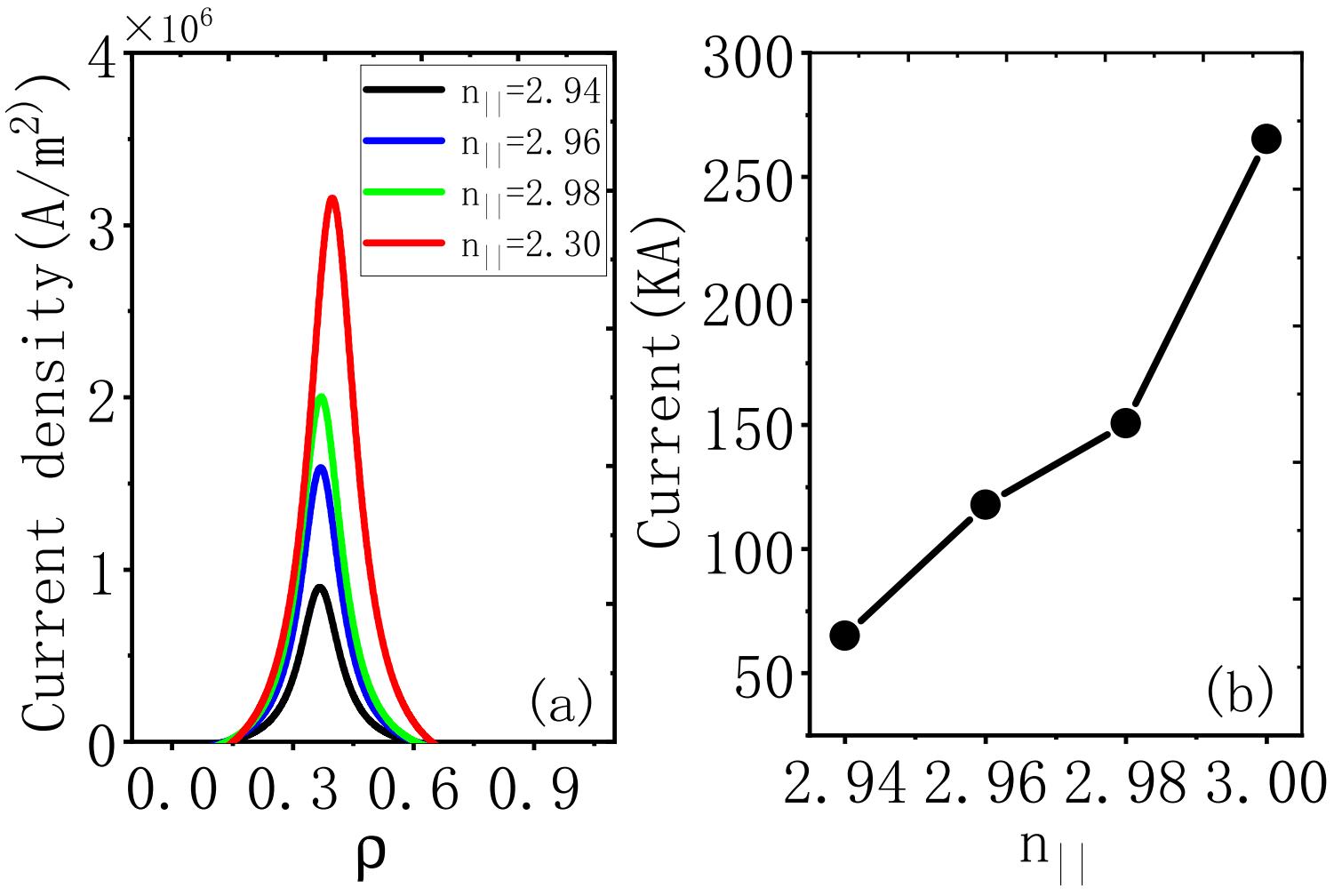}
\caption{Helicon wave of 170~MHz driven current versus $n_{||}$ with $n_{e0}=0.8\times 10^{18}~\mathrm{m}^{-3}$}
\label{fig8}
\end{figure}

Fig.~\ref{fig7} shows the change of driven current with $n_{||}$ with the helicon wave frequency of 150~MHz and 170~MHz. The central electron density is $2.4\times 10^{18}~\mathrm{m}^{-3}$ here. It can be seen that when the incident wave frequency is 150~MHz and 170~MHz, there is a specific positive correlation between the driven current and $n_{||}$ within the range of interest ($n_{||} = 3.2$--4.5). The Figure also depicts that the wave drive capability at lower frequencies is relatively lower than at higher frequencies. In order to study the influence of  $n_{||}$ on the driving current profile in more detail, we scanned the parallel refractive index of the 170~MHz helicon wave in a narrow range. The results are shown in Fig.~\ref{fig8}. It is worth noting that the central electron density is $0.4\times 10^{18}~\mathrm{m}^{-3}$ in this case. It indicates that within the concerned parameter range, there is a particular proportional relationship between the driven current and $n_{||}$. Moreover, this Figure clearly show that as $n_{||}$ increases, the driving distribution becomes wider.

\subsection{Influence of SOL region on helicons current drive}
\label{sec3.3}
C. Lau et al. investigated the helicon wave current drive in DIII-D and ITER tokamaks with the full-wave code AORSA. They focused on the simulations with the SOL and showed that high SOL density could result in significant helicon wave power loss ($\sim$10--20\%) in the SOL.~\cite{6} Therefore, the influence of SOL needs to be considered for a more accurate calculation of the helicon drive current.

\begin{figure}[!ht]
\centering
\includegraphics[width=0.8\columnwidth]{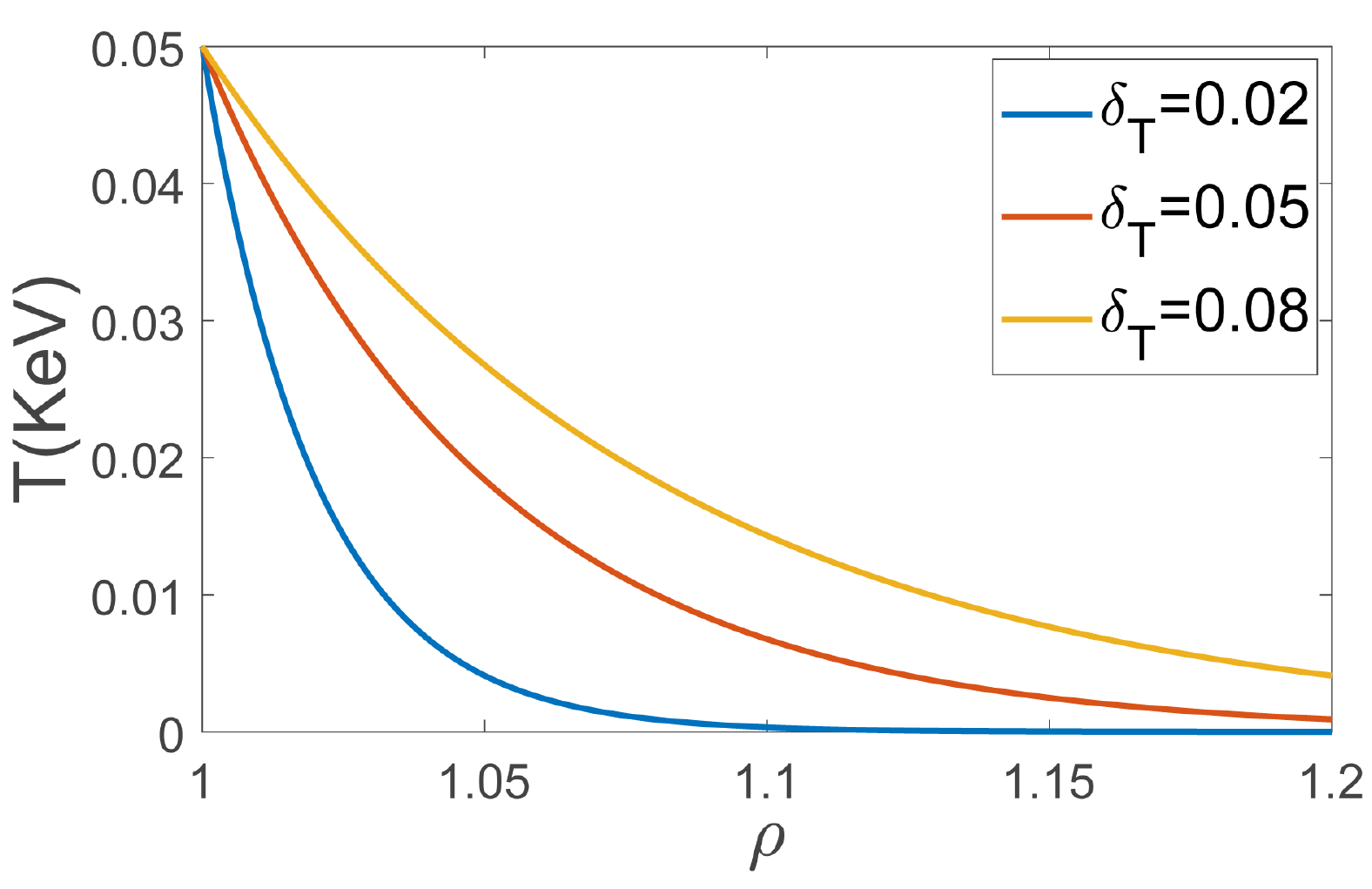}
\caption{Temperature profiles in the SOL region at different $\delta_T$}
\label{fig9}
\end{figure}

As mentioned earlier, we changed the temperature and density distribution of the plasma in the SOL region by altering $\delta_{T(n)}$ as per Eq. 3. For example, the temperature distribution outside the LCFS, at different $\delta_T$ is shown in Fig.~\ref{fig9}. As the $\delta_{T(n)}$ gets smaller and smaller, the temperature or density distribution outside the LCFS decreases faster and faster.
\begin{figure}[!ht]
\centering
\includegraphics[width=0.8\columnwidth]{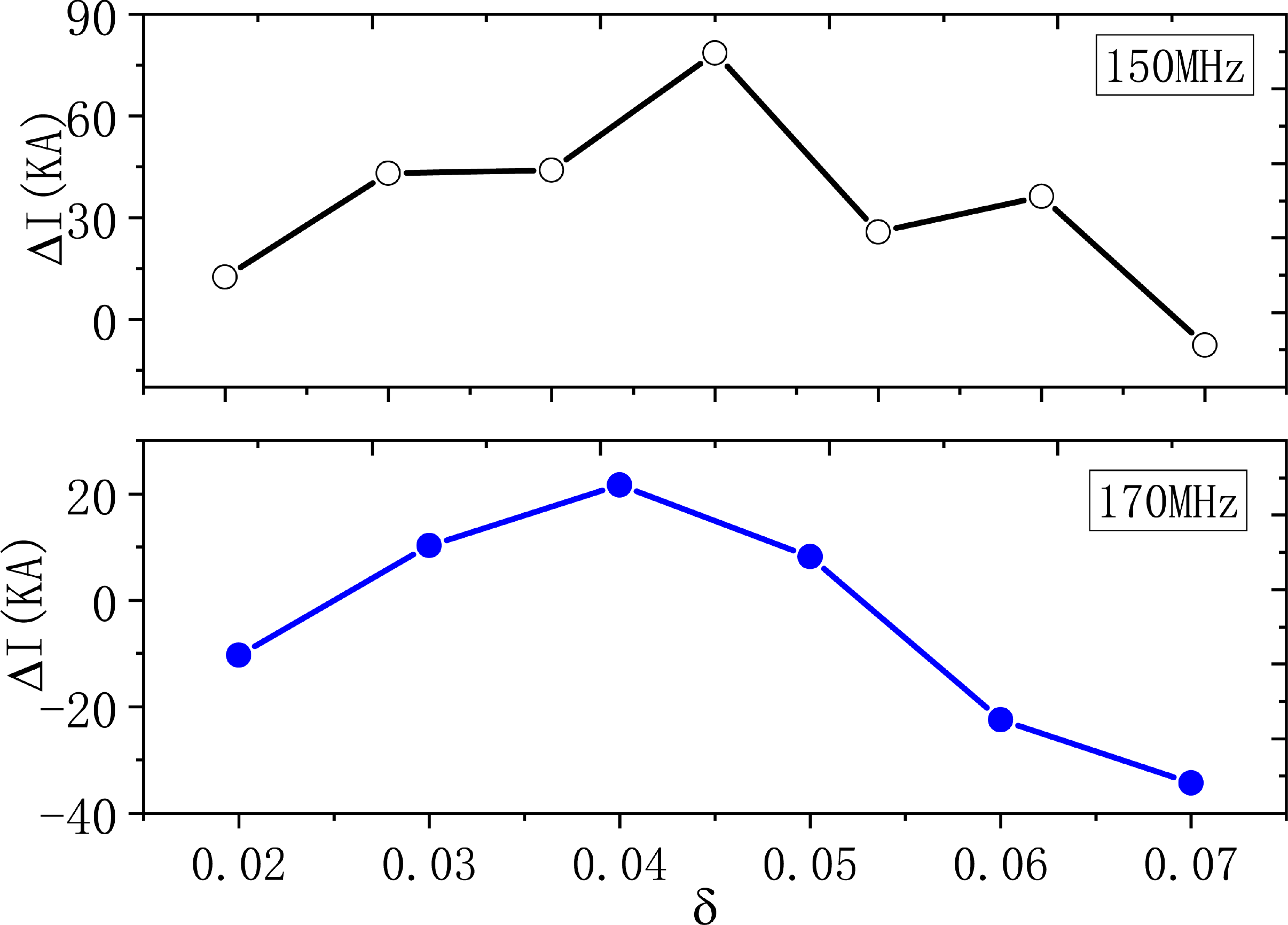}
\caption{The relationship between the change of the helicon driven current and the parameter $\delta_{T(n)}$ of the SOL region}
\label{fig10}
\end{figure}

Fig.~\ref{fig10} shows the difference between the driven current with and without the SOL region as a function of $\delta_{T(n)}$. The top panel corresponds to the 150~MHz case, and the bottom panel to the 170~MHz case. It can be seen that, generally, $\Delta I$ is positive, indicating that in most parameter intervals, the SOL region reduces the current driven by the helicon waves. Specifically, for 150~MHz, when $\delta = 0.045$, the driven current decreases the most, up to 70~KA. If this value continues to increase, $\Delta I$ will become negative. Considering the SOL region at this time, the driven current can be increased. However, for 150~MHz, the increase in the magnitude of the current is less. For 170~MHz, the trend of current change is the same as that of 150~MHz. The difference is that the decreasing current amplitude is low (e.g., $\delta  = 0.04$), and the increasing amplitude is higher (e.g., $\delta  = 0.07$). It is to be noted that when the SOL region is not considered, the currents driven by 150~MHz and 170~MHz helicon waves are 134~KA and 123~KA, respectively, which is consistent with the Fig.~\ref{fig7}.

\section{Summary and discussion}
\label{sec4}
With the EXL-50 tokamak discharge parameter, it is found that when the helicon wave frequency is 300~MHz$\sim$380~MHz, with the increment of $n_{||}$, the driven current first increases, then decreases, and peaks at around $n_{||}$ = 4.0. Similar to the previous calculations, the driven current of helicon waves shows a relatively stable upward trend with the increment of plasma temperature. With the changes in the plasma density, the driven current decreases. The helicon wave can drive a considerable current when the plasma density is extremely low. For example, when the density is $1.0\times 10^{18}~\mathrm{m}^{-3}$, a current of about 500~KA can be driven with a 1~MW helicon wave.

For the discharge of EXL-50 tokamak, we conducted a more detailed study on the helicon wave current drive at 150~MHz and 170~MHz. It was found that the driven current has a relatively stable upward trend at lower frequencies with the increase in plasma temperature. At the same time, we note that at these two helicon wave frequencies (150~MHz and 170~MHz), a higher plasma density will be deleterious to generating a driven current. Meanwhile, it is also found that the driven current at lower frequencies is relatively lower than those at higher frequencies. In the range of the considered $n_{||}$ values, the driven current has a positive proportional relationship with $n_{||}$. Moreover, the driven current profile gets more expansive with the increase of $n_{||}$.

Finally, we investigated the impact of the SOL region on the wave current drive. We found that after considering the SOL region, the current drive of the helical wave is heavily dependent on the $\delta $ parameter, which affects the density and temperature profile in the SOL region. With various $\delta $ values, considering the SOL region, the driven current can become larger or smaller. In general, when $\delta =0.04$, considering the SOL region, the wave-driven current is reduced, and the reduction reaches the maximum value.

Our study will be an essential reference for the engineering design of helicon wave systems, experiments of EXL-50 tokamak, and the other spherical tokamaks. Future work includes the quasi-linear study and the parameter decay investigation of helicon waves in spherical tokamaks.

\section*{Acknowledgements}
The authors gratefully acknowledge Prof. Zhihong Lin (UCI) for fruitful discussions. This work is supported by NSFC (Nos. 11905109 and 11947238), the National Key R\&D Program of China (Nos. 2018YFE0303102) , National magnetic confinement fusion energy development research project (Nos. 2022 YFE03070003), Natural Science Foundation of Hunan Province (Nos. 2020J J4515), Key projects of Hunan Provincial Department of Education (Nos. 20A432), and the Government Sponsored Study Abroad Program of the Chinese Scholarship Council (CSC) (Nos.  202108430056).








\end{document}